\documentclass[twocolumn,10pt]{asme2e}

\usepackage{graphicx}
\usepackage{amsmath}

\confshortname{MSEC 2016}
\conffullname{the ASME 2016 Manufacturing Science and Engineering Conference}

\confdate{June 27-July 1}
\confyear{2016}
\confcity{Blacksburg}
\confcountry{USA}

\papernum{MSEC2016-8849}

\title{DRAFT: Evaluation of Cellular Solids Derived from Triply Periodic Minimal Surfaces}

\author{Daniel Cellucci\thanks{Address all correspondence to this author.}\\
    \affiliation{
	Space Systems Design Studio\\
	Department of Mechanical and Aerospace Engineering\\
	Cornell University\\
	Ithaca, NY 14850\\
    Email: dwc238@cornell.edu
    }	
}

\author{Kenneth C. Cheung\\ 
    \affiliation{Coded Structures Lab\\
	NASA Ames Research Center\\
	Moffett Field, CA 94035\\
	Email: kenny@nasa.gov
    }
}

\begin{document}

\maketitle    

\begin{abstract}
{\it Cellular solids are a class of materials that have many interesting engineering applications, including ultralight structural materials~\cite{cheung2013reversibly}. The traditional method for analyzing these solids uses convex uniform polyhedral honeycombs to represent the geometry of the material~\cite{gibson1997cellular}, and this approach has carried over into the design of digital cellular solids~\cite{cheung2012digital}. However, the use of such honeycomb-derived lattices makes the problem of decomposing a three-dimensional lattice into a library of two-dimensional parts non-trivial. We introduce a method for generating periodic frameworks from Triply Periodic Minimal Surfaces (TPMS), which result in geometries that are easier to decompose into digital parts. Additionally, we perform multiscale analysis of two cellular solids generated from two TPMS, the P- and D-Schwarz, and two cellular solids, the Kelvin and Octet honeycombs. We show that the simulated behavior of these TMPS-derived structures shows the expected modulus of the cellular solid scaling linearly with relative density, and matches the behavior of the octet truss.}
\end{abstract}

\begin{nomenclature}
\entry{$\rho^*$}{density of the cellular solid}
\entry{$\rho$}{density of the constituent material}
\entry{$\bar{\rho}$}{relative density between the cellular and constituent solid}
\entry{$\phi_t$}{family parameter relating a TPMS and its conjugate}
\entry{$K_{uc}$}{Stiffness matrix of a single unit cell}
\entry{$K_{mat}$}{Effective material stiffness matrix}
\entry{$E_s$}{Constituent material Young's modulus}
\end{nomenclature}

\section*{INTRODUCTION}

Cellular solids are a class of materials composed of an interconnected network of struts and plates, which form the edges and faces of cells that tesselate three-dimensional space~\cite{Ashby:2006aa}. The geometry of these cells can be represented by the constituent polyhedra in a convex uniform honeycomb. There are 28 such honeycombs~\cite{grunbaum1994uniform}, and common examples of the derived frameworks include the Kelvin lattice, the Octet lattice, and the Cuboctahedral lattice.

Cellular solids are found in many forms in nature, and have been applied in many engineering problems, including the construction of ultralight structural materials~\cite{gibson1997cellular,cheung2013reversibly}. The most important characteristic of these solids is the relative density $\rho^*/\rho$, which relates many of the global properties of the cellular solid to the bulk material from which the solid is derived~\cite{Ashby:2006aa}. Some of these properties, modulus and ultimate strength, depend on the properties of the geometry chosen to represent the cellular solid. One such property is the coordination, or the number of struts that meet at a given node.

For instance, in geometries where there is sufficient coordination between the frame elements that loads are transferred axially, such as in the Octet lattice, the foam is said to be “stretch-dominated” and the Young's modulus of the material is directly proportional to the relative density. In geometries where there is insufficient coordination, such as in the Kelvin lattice, loads are transmitted between the struts through bending, so the foam is said to be “bending-dominated”, and the Young's modulus of the materials is quadratically proportional to the relative density~\cite{Ashby:2006aa}. Furthermore, the range of behaviors that can be reached by varying the geometry is not limited to these two; coordinated-buckling modes such as those in the Cuboctahedral lattice can impart a mixed stretch-bending behavior to load transfer, and therefore the resulting Young's modulus is proportional to the relative density to the $3/2$ power~\cite{cheung2013reversibly}. 

Recent work in digital cellular solids introduced a decomposition of the cuboctahedral lattice into a single, two-dimensional part, which enabled the reversibly assembled into structures with, at the time, unprecedented strength-for-weight~\cite{cheung2013reversibly}. This work showcased many desirable traits to the scalable assembly of cellular solids, including centimeter-scale construction, repairability, and mass-production of parts from high-performance materials~\cite{cheung2012digital}. Subsequent work has demonstrated further capabilities of this approach, including robotic traversal\cite{MOJO} and meter-scale structures~\cite{yosemite} using a three-part strut-node construction.

However, in both digital and conventional cellular solids, the geometry that is conventionally used for modeling a manufactured foam or as the basis for a digital decomposition is one that has been generated from a convex uniform honeycomb. While it is clearly justified for studying the theoretical behavior of stretch- versus bending- dominated foams, the structures derived from convex uniform honeycombs have nodes (points where the struts of the cellular solid meet) that are three-dimensional. This therefore makes the problem of decomposing such a lattice into two-dimensional parts, and therefore access to many of the advantages of digital cellular solids, non-trivial. 

\subsection*{Triply Periodic Minimal Surfaces}

An alternative approach to using convex uniform polyhedra in the design of digital cellular solids is to find periodic frameworks whose node geometries are simpler, but whose performance is comparable to existing examples. 

Triply periodic minimal surfaces (TPMS) are embedded surfaces that are translationally invariant in three orthogonal directions~\cite{karcher1996construction}. These objects have many advantageous properties as a generator of periodic frameworks for cellular solids. Since they are non self-intersecting surfaces, all lines on the surface will meet on a locally flat plane. If these lines are taken as the struts of a periodic framework, and the nodes as the intersections between the lines, then the complexity of these nodes is greatly reduced compared to equivalent-connectivity structure generated from the edges and vertices of a convex uniform polyhedral honeycomb. Additionally, since TPMS are triply periodic, a framework found for a single unit cell will also be periodic and therefore an appropriate geometry for a periodic framework.

This paper seeks to apply the analytical tools originally developed for studying cellular solids to frameworks generated from triply periodic minimal surfaces. It discusses a method for deriving a periodic framework from a TPMS, and then studies the theoretical static performance of this framework compared to conventional cellular solids.

\section*{Methodology}

A TPMS can be represented in a variety of ways. When first discovered, examples were generated using the Enneper-Weierstrau\"s formulas~\cite{karcher1996construction}, which parametrize the minimal surface in the complex domain. The complexity of this representation limited its usefulness, and even today only a few such representations of various TPMS are known. Later, Shoen used interpenetrating skeletal graphs with specific crystallographic symmetry to visualize a wide variety of TPMS~\cite{schoen1970infinite}. However, it wasn’t until Karcher formalized the descriptions of Shoen's TPMS using the conjugate surface method that they gained traction in the mathematical community~\cite{karcher1989triply}.

The conjugate surface method takes advantage of a fundamental observation of minimal surfaces- every minimal surface belongs to an associate or Bonnet family, where a single family parameter $\phi_t \in \{0,2\pi\}$ can produce a continuous range of surfaces~\cite{karcher1989triply}. In addition to generating an infinite variety of minimal surfaces through variation of the family parameter, this method guarantees a minimal patch that can be converted to a periodic framework, as shown in the section below.

In the natural sciences, implicit approximations of the TPMS have been used instead of the explicit forms~\cite{gandy2001nodal}, in order to save computational overhead. In particular, the use of periodic nodal surfaces have satisfactorily reproduced the most famous TPMS without extending beyond the leading term of the series, and higher order approximations have resulted in even better fidelity. Subsequent calculations performed on TPMS in this paper will use the periodic nodal surface approximations instead of the more complex parameterizations.

\subsection*{Finding the Periodic Frameworks}

The periodic frameworks that can be derived from a given TPMS are the embedded straight lines of the surface. A general method for finding these lines will allow the infinite variety of TPMS to act as generators for periodic frameworks.

An analytical approach for finding such lines is to use the conjugate surface method to solve the free boundary problem for the asymmetric unit of the conjugate TPMS. The resulting decomposition of the polyhedral asymmetric unit into the polygon boundary of the desired surface produces the set of straight lines that can then be taken as the primitive unit of the periodic framework~\cite{karcher1996construction}. 

For example, to find the periodic framework for the P-Schwarz surface, we would start with the asymmetric unit of its conjugate surface, the D-Schwarz. This unit is a tetragonal disphenoid, or an isohedral tetrahedron with isosceles triangle faces~\cite{gandyexactD}. This unit then forms the set of planes that the free boundary will be solved on by constructing a conjugate polygonal contour, through a process described by Karcher and Polthier~\cite{karcher1996construction}. This contour forms the boundary of the minimal patch for the P-Schwarz, which can be transformed through the symmetries of the P-Schwarz’ point group to produce the surface~\cite{karcher1989triply}. The lines which form the polygonal boundary of this minimal patch are the embedded periodic framework of the surface. 

This method, however, only works for simpler TPMS ~\cite{gandy2001nodal}, since more complex structures introduce unknown degrees of freedom that must be balanced to avoid self-intersections. There is no general theorem to solve this so-called period problem, and so the applicability of this method is limited~\cite{karcher1996construction}. 

A numerical approach can also be employed to find the straight lines. This approach has the advantages of being general- all that is required is an implicit approximation of the surface. A downside of the approach, however, is that the existence of the straight lines cannot be ascertained using this method.

\begin{enumerate}
\item Begin with a TPMS with an implicit approximation $f(\vec{x}) = 0$, defined inside a unit cell $\vec{x} \in \{0,1\}$
\item Choose two faces of the unit cell, and define a curve on each of these faces, $C_1$and $C_2$, which represent the intersection between the TPMS and the chosen unit cell face
\item For each point $\vec{p}_1$ on $C_1$, we define a vector $\vec{v}$ between that point and every point $\vec{p}_2$ on $C_2$, $\vec{v} = \vec{p}_2-\vec{p}_1$
\item If $f(\vec{p}_1-t\vec{v}) = 0$ for all $t \in \{0,1\}$, then the pair $(\vec{p}_1,\vec{p}_2)$ represents an embedded line on the surface. 
\item Repeat for all pairs of surfaces
\end{enumerate}

\begin{figure}[t]
\begin{center}
\includegraphics[width=0.4\textwidth]{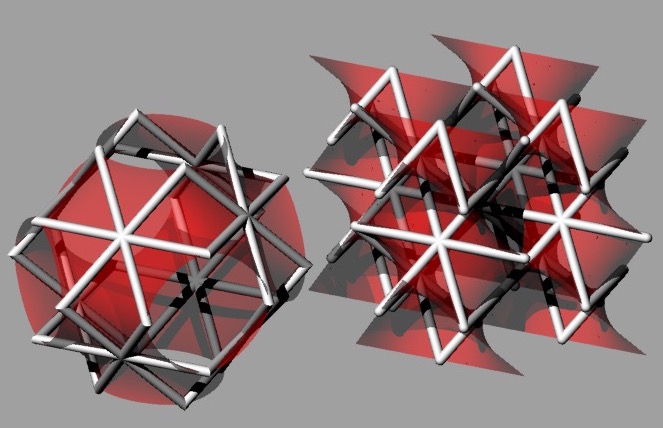}
\end{center}
\caption{THE PERIODIC FRAMEWORK FOUND FOR THE P-SCHWARZ (LEFT) AND D-SCHWARZ (RIGHT) SURFACES}
\label{figure_frame} 
\end{figure}

We apply this approach to the P- and D-Schwarz surfaces, using a numerical method that splits the boundary curves of each of the sides into 1000 points, and assesses 10 equally-spaced points along the length of each test line to see if it is embedded. The set of lines is shown in Figure~\ref{figure_frame}. Furthermore, inspection of these lines shows additional intersections that aren’t apparent from the numerical algorithm, but are explicitly defined in the polygonal contours found with the analytical approach.

\section*{Analysis}

Inspection of the frameworks derived from the P- and D-Schwarz surfaces shows that they are composed of four sets of interpenetrated isogrid lattices whose pitch is four times the length of a single strut. These lattices intersect at the midpoints of the isogrid elements in the D-Schwarz geometry, and at the first and third quarter of the isogrid elements in the P-Schwarz geometry. This results in a different node geometry for the two frameworks- 2/5 of the nodes in the P-Schwarz framework have connectivity 6 and the other 3/5 connectivity 4, while the D-Schwarz has only nodes with connectivity 6.

\subsection*{Theoretical behavior}

We applied Vigliotti and Pasini's multiscale stiffness analysis to the P- and D-Schwarz frameworks to assess the mechanical response of the bulk cellular solid under different loading conditions~\cite{vigliotti2012stiffness}. This technique applies periodic boundary conditions to derive properties of an infinite lattice using the stiffness matrix of a single unit cell $K_{uc}$. In addition, we also applied this analysis to the Kelvin and Octet lattices, whose mechanical properties are already well known, in order to ensure that the results were consistent.

Because of the cubic symmetry of all four of these lattices, the macroscopic stiffness matrix can be written as 

\begin{equation}
K_{mat} = \left[ \begin{array}{c c c c c c}
\alpha &  \beta &  \beta & 		0 &		 0 & 	  0 \\
 \beta & \alpha &  \beta & 		0 &		 0 & 	  0 \\
 \beta &  \beta & \alpha & 		0 &		 0 & 	  0 \\
	 0 & 	  0 &	   0 & \gamma &		 0 & 	  0 \\
	 0 & 	  0 &	   0 & 		0 &	\gamma & 	  0 \\
	 0 & 	  0 &	   0 &		0 &		 0 & \gamma 
\end{array} \right]
\end{equation}
 
The inverse of $K_{mat}$ relate the stresses in the cubic coordinate system to the strains.

\begin{equation}
\left(\begin{array}{c}
\epsilon_x\\
\epsilon_y\\
\epsilon_z\\
\epsilon_{yz}\\
\epsilon_{xz}\\
\epsilon_{xy}\end{array} \right) = 
\left( \begin{array}{c c c c c c}
  s_1 & -s_2 & -s_2 & 	0 &	  0 & 	0 \\
 -s_2 &  s_1 & -s_2 & 	0 &	  0 & 	0 \\
 -s_2 & -s_2 &  s_1 & 	0 &	  0 & 	0 \\
	0 &	   0 &	  0 & s_3 &	  0 & 	0 \\
	0 &	   0 &	  0 &   0 &	s_3 &   0 \\
	0 &	   0 &	  0 &	0 &	  0 & s_3 
\end{array} \right)
\left(\begin{array}{c}
\sigma_x\\
\sigma_y\\
\sigma_z\\
\sigma_{yz}\\
\sigma_{xz}\\
\sigma_{xy}\end{array} \right)
\end{equation}

$K_{mat}^{-1}$, otherwise known as the compliance matrix, can be normalized for the constituent material stiffness $E_s$ and the relative density of the cellular solid $(\bar{\rho})^n$, to produce a function

\begin{equation}
\label{relation}
\frac{1}{s_i} = k_i \bar{\rho}^{n_i} E_s
\end{equation}

Where the values of $n$ and $k$ depend on the geometry.

\begin{table}[]
\centering
\caption{RELATIVE DENSITIES FOR THE FOUR LATTICES EXAMINED, AS A FUNCTION OF THE STRUT LENGTH $l$ AND CROSS-SECTIONAL AREA $a$}
\label{tab:relden}
\begin{tabular}{c|c}
Lattice   & $\bar{\rho} $                       \\
\hline
Kelvin    & $\frac{3}{2\sqrt{2}}\frac{a}{l^2}$ \\
D-Schwarz & $3\sqrt{2}\frac{a}{l^2}$            \\
P-Schwarz & $\frac{3}{2\sqrt{2}}\frac{a}{l^2} $ \\
Octet     & $6\sqrt{2}\frac{a}{l^2}$           
\end{tabular}
\end{table}

\begin{figure*}[t]
\begin{center}
\includegraphics[width=1.0\textwidth]{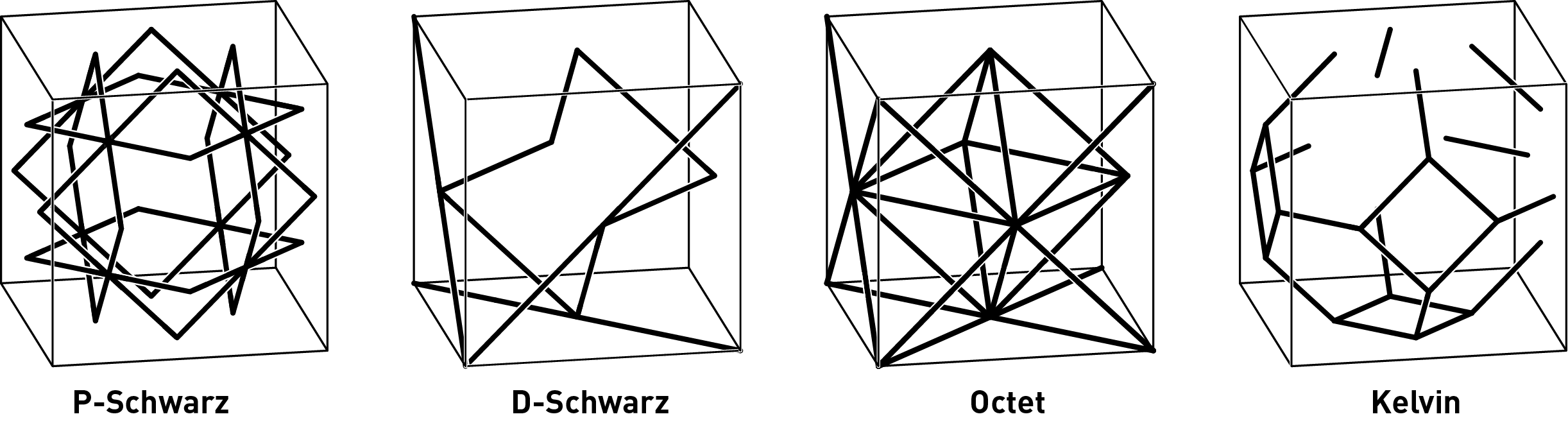}
\end{center}
\caption{UNIT CELLS FOR THE FOUR LATTICES STUDIED. DUE TO PERIODIC BOUNDARY CONDITIONS, ONLY THE FRAMES WHICH CONTRIBUTE TO $K_{uc}$ ARE SHOWN.}
\label{fig_lattices} 
\end{figure*}

The unit cells used to generate $K_{uc}$ are shown in Figure~\ref{fig_lattices} for the four different lattices. The frames of each lattice were modeled as Timoshenko Beams, with circular cross-sections. $K_{uc}$ was then transformed into the macroscopic stiffness $K_{\epsilon}$. The eigenvalues of $K_{\epsilon}$ provided the values for $\alpha$, $\beta$, and $\gamma$, which allowed the construction of $K_{mat}$. Finally, $K_{mat}$ was inverted to provide the the compliance matrix and therefore the expected macroscopic performance.

For all of these lattices, the strut properties for $K_{uc}$ were calculated from three different relative densities- 0.1, 0.01, and 0.001- using the equations for relative density described in Table~\ref{tab:relden}. This allowed the calculation of $k_i$ and $n_i$ from Equation~\ref{relation}. For the Kelvin lattice, these values were found to be

\begin{subequations}
\begin{align}
        \frac{1}{s_1} &= \frac{1}{1.681}\bar{\rho}^2 E_s,\\
        \frac{1}{s_2} &= \frac{1}{0.826}\bar{\rho}^2 E_s,\\
        \frac{1}{s_3} &= \frac{1}{5.095}\bar{\rho}^2 E_s.
\end{align}
\end{subequations}

The value of $s_1$ shows good agreement with previous analytical work ~\cite{warren1997linear}.

The application of this technique to the D-Schwarz, P-Schwarz and Octet produced identical coefficients for the major compliances.

\begin{subequations}
\begin{align}
        \frac{1}{s_1} &= \frac{1}{9}\bar{\rho} E_s,\\
        \frac{1}{s_2} &= \frac{1}{3}\bar{\rho} E_s,\\
        \frac{1}{s_3} &= \frac{1}{12}\bar{\rho} E_s
\end{align}
\end{subequations}

These values agree with previous theoretical work for Octet~\cite{deshpande2001effective}, and suggest that the macroscopic material properties of P-Schwarz and D-Schwarz not only scale linearly with relative density, but that the coefficients with which they scale are equivalent to Octet. 

Furthermore, previous study of the linear elastic behavior of the Octet lattice showed that, in the (111) direction (otherwise known as the space diagonal) of the unit cell, the stiffness scaled with $E_{111} = (\bar{\rho}/5)E_s$, which is the maximum value of the Young's modulus of the Octet lattice for all orientations~\cite{deshpande2001effective}. Since this modulus was found by performing a coordinate transfer on the compliance matrix, we can expect the same performance from the P-Schwarz and D-Schwarz as well.

\section*{Discussion}

The simulated behavior of both structures indicates that their response to an applied load is equivalent that of the far more complex Octet truss, which has coordination 12 and is much harder to decompose into two-dimensional parts.

\subsection*{A Digital Design}
\begin{figure}[h!]
\begin{center}
\includegraphics[width=0.5\textwidth]{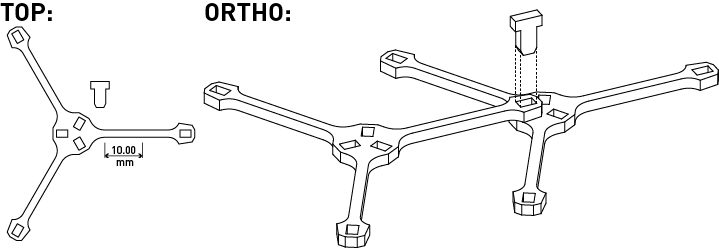}
\end{center}
\caption{TWO-DIMENSIONAL PART DESIGN FOR THE D-SCHWARZ USING A HEXAGONAL STRUT PLATE WITH A PRESS-FIT LOCKING PIN.}
\label{figure_parts} 
\end{figure}

While finding the periodic framework for a given TPMS is decidedly less trivial than finding a framework for a given convex uniform polyhedron honeycomb, the reverse is true for generating a digital decomposition of that framework. Because the node geometry of the TPMS lattices are confined to a two-dimensional plane, the decomposition is straightforward, and one such decomposition is shown in Figure~\ref{figure_parts}. 

\begin{figure}[h!]
\begin{center}
\includegraphics[width=0.5\textwidth]{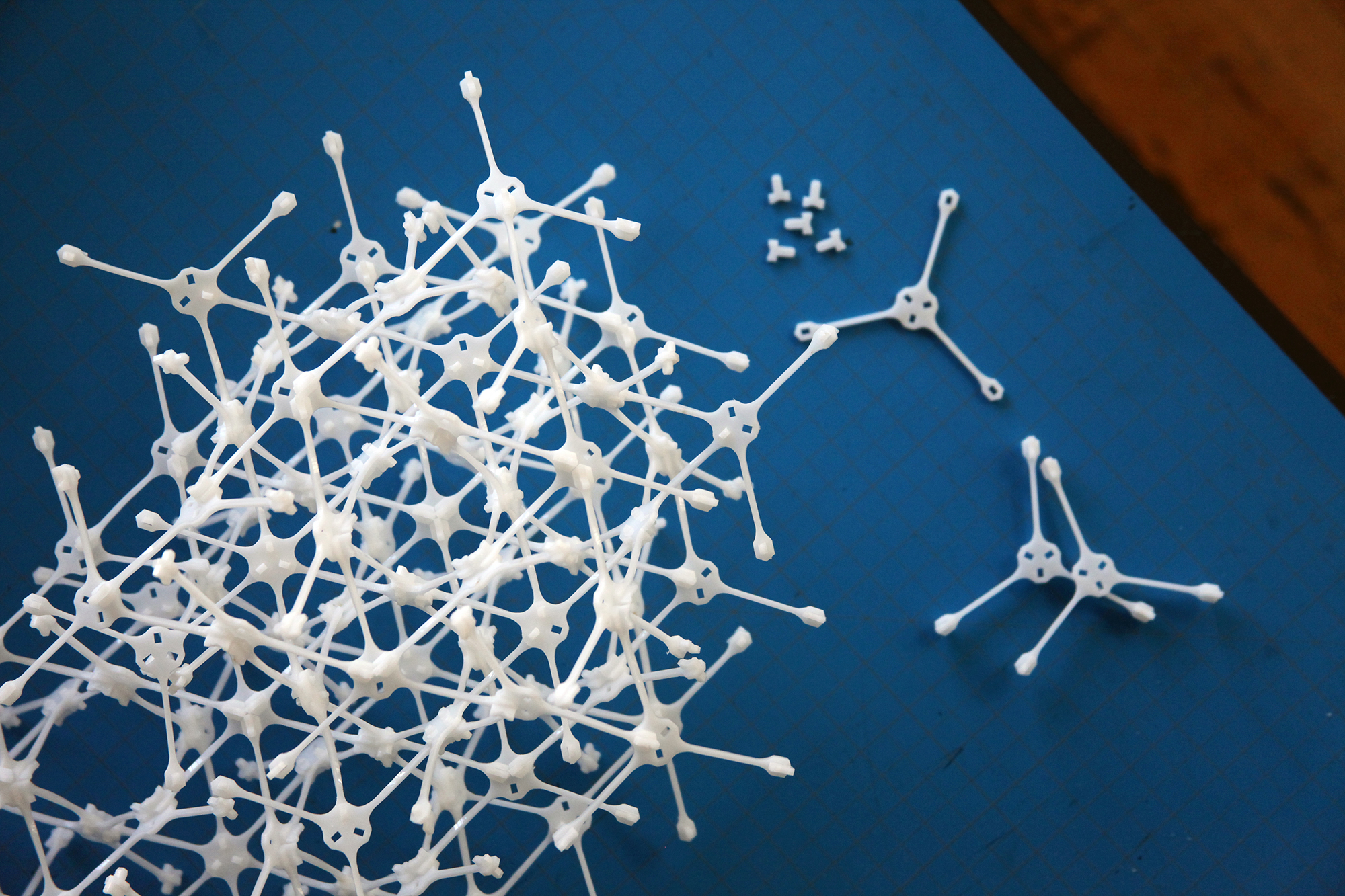}
\end{center}
\caption{A PHYSICAL MODEL OF THE D-SCHWARZ FRAMEWORK CONSTRUCTED FROM A LIBRARY OF TWO PARTS. THE PARTS ARE MADE OUT OF LASER-CUT DELRIN AND PRODUCE A LATTICE WITH A PITCH OF 3 INCHES.}
\label{figure_model} 
\end{figure}

Using this design, the D-Schwarz can be constructed out of a library of two parts: a part with three-fold symmetry, three arms and three slots, and pins for attaching the arms to a neighboring part's slot. The P-Schwarz can be constructed from three parts consisting of the two used for the D-Schwarz, and an additional adapter plate with four-fold symmetry consisting of two arms and two slots. Figure~\ref{figure_model} shows a prototype D-Schwarz structure constructed out of laser-cut $3/32"$ Acetal Delrin. 

An astute observer will note that, while the nodal geometry is locally flat, a two-dimensional part that composes TMPS framework will have to twist $54^{\circ}$ with the P-Schwarz and $70^{\circ}$ for the D-Schwarz. One solution involves constructing parts with this twist pre-defined, so that there is no pre-stress imposed on the structure. Future experimental work will be devoted to seeing what effect the twist has on the performance of the structure. 

\begin{acknowledgment}
This work was funded by the NASA Space Technology Research Fellowship.
Additional thanks go to NASA Ames Research Center for providing the facilities and the environment that made this work possible.
Finally, the authors appreciate the guidance of Dr. Theodore Shifrin, in identifying a path toward finding embedded lines on the surface of a TPMS.
\end{acknowledgment}

\bibliographystyle{asmems4}

%
% ...

\bibliography{sources}

\end{document}